\documentclass[superscriptaddress,aps,prd,floatfix,nofootinbib,bibnotes]{revtex4-1}
\usepackage[colorlinks=true,linkcolor=black,citecolor=blue, urlcolor=blue,bookmarks=false]{hyperref}
\hypersetup{breaklinks=true}
\usepackage[utf8]{inputenc}
\usepackage{natbib}
\usepackage{array}
\usepackage{epsfig}
\usepackage{amsmath}
\usepackage{amssymb}
\usepackage{multirow}
\usepackage{graphicx,subcaption}
\usepackage{enumitem}
\graphicspath{{./fig/}}
\usepackage[normalem]{ulem}  
\usepackage[dvips]{color} 

\renewcommand{\sout}{\bgroup \color{red} \ULdepth=-.5ex \ULset}

\begin{document}

\title{Chiral symmetry breaking and the masses of hadrons: a review}

\author{Su Houng Lee}
\email{suhoung@yonsei.ac.kr}
\affiliation{Department of Physics and Institute of Physics and Applied Physics, Yonsei University, Seoul 03722, Korea}

\begin{abstract}
The masses of hadrons in the vacuum, where the chiral symmetry is restored, and in the medium are in general different even when the changes in the order parameters of chiral symmetry are the same.  Here, we first discuss the relation between the hadron masses and the chiral symmetry breaking in approaches based on operator product expansion (OPE). 
We then discuss what additional changes occur to the hadron masses when going from the chiral symmetry restored vacuum to nuclear medium and/or finite temperature. 
The work will highlight how we can identify the effects of chiral symmetry restoration from experimental observations.  
\end{abstract}

\maketitle


\section{Introduction}

We typically express the masses of composite particles as the sum of their constituents and small binding energies.  Such decomposition is not possible in quantum chromodynamics (QCD) as 
confinement and chiral symmetry breaking are intricately related to in generating the hadron mass. 
Yet, it is known that chiral symmetry breaking is an important ingredient, so isolating its effects would provide an important cornerstone in understanding the origin of the hadron mass\cite{Hatsuda:1985eb,Brown:1991kk,Hatsuda:1991ez,Leupold:2009kz}.  

In the original application of the QCD sum rules approaches to the vector meson masses in the nuclear medoum\cite{Hatsuda:1991ez},
the meson masses were found to change due to the changes of the light 4-quark condensates for the $\rho$, $\omega$ mesons and the strange quark condensates for the $\phi$ meson.  In the case of $J/\psi$ in the medium, the changes occur due to the change of gluon condensate, which is related to the scale-breaking effects through trace anomaly.  
Since then, experiments have been performed worldwide to observe mass shifts of hadrons from heavy ion collisions and nuclear target experiments, which are expected to reflect finite temperature or density configurations, respectively \cite{Hayano:2008vn,Ichikawa:2018woh}.

Experimental efforts were concentrated on measuring the electromagnetic signals from heavy ion collisions and nuclear target experiments.  
While experiments trying to measure dilepton spectra from heavy ion collisions are still going on and planned in the future, these signals have many sources throughout the evolution of the system. Furthermore, information about chiral partners is hidden in the continuum. On the other hand, nuclear target experiments have advantages: the target nuclear density remains almost constant, and even at these densities, the order parameter is known to decrease substantially around 30$\%$\cite{Lacour:2010ci,Meissner:2001gz}.
Also, signals from hadronic decays are viable as long as one, focuses on small-width mesons. Furthermore, one can combine excitation function measurements and the  transparency 
ratios~\cite{Metag:2017yuh} to estimate the mass and width changes simultaneously.
But in order to identify the effect of chiral symmetry restoration, it seems essential to first look at the chiral partners and measure their mass differences.  Such measurement will establish the cornerstone for the relation between mass shift and chiral symmetry restoration.  One should then measure the individual masses.  

In this review, we will summarize what happens to hadrons in medium from the perspective of operator product expansion (OPE).  We will 
first show how to isolate the chiral symmetry-breaking effects in the hadron masses in the OPE perspective.  The method will be applied to the vector mesons and then to other hadrons. Through this method, we will be able to isolate the effects of chiral symmetry breaking and thus calculate the mass of hadrons in the chiral-symmetry restored vacuum.  
Here, the chiral-symmetry restored vacuum means the QCD configuration where the chiral-symmetry breaking effects are taken away while other non-perturbative effects are left intact.  Such a configuration can be probed in the OPE formalism by taking all the quark operators proportional to the chiral order parameter to be zero while keeping all other quark or gluon operators intact.  From a lattice point of view, one can also probe such a configuration by taking away all the quark zero modes when calculating hadron properties.
We will then discuss what other effects will emerge when we place hadrons in a medium\cite{Kim:2020zae,Kim:2021xyp}: these effects do not contribute when one looks at the differences between the chiral partners.  Finally, we introduce the mass of $K_1$
and $K^*$ as a suitable example for chiral partners that can be realistically measured in a nuclear target experiment\cite{Song:2018plu,Lee:2019tvt}. 

\section{Chiral order parameter}

In this section, we discuss how to isolate the chiral symmetry-breaking part of any quark operator\cite{Kim:2020zae,Kim:2021xyp}: the breaking part is the chiral order parameter. 
In any expectation value, the quark part will involve the quark propagator $S_q(x,0) \equiv S_q(x,y)|_{y=0}$, which can be written into a part that is symmetric and anti-symmetric under chiral rotation: this can be accomplished by adding and subtracting its chiral partner.  The respective parts are also denoted as chiral even $(S)$  and odd $(B)$  components.
\begin{eqnarray}
S_q(x,0) &=&  \bigg( S_q^B(x,0)+ S_q^S(x,0) \bigg),  \label{prop0} \\
S_q^B(x,0) &=& \frac{1}{2} \bigg( S_q(x,0)-i \gamma^5 S_q(x,0)i\gamma^5 \bigg),  \label{propb}\\
S_q^S(x,0) &=& \frac{1}{2} \bigg( S_q(x,0)+i \gamma^5 S_q(x,0)i\gamma^5 \bigg). 
\label{props}
\end{eqnarray}
The separation can also be understood as making a specific chiral transformation to the quark operator $q\rightarrow \gamma^5 q$ so that $S_q(x,y) \rightarrow i \gamma^5 S_q(x,y) i\gamma^5$ leading to the symmetric and breaking part as given in Eq.\eqref{props} and Eq.\eqref{propb}, respectively.

The usual chiral condensate is the dimension 3 two-quark operator.  Here, because of the trace or due to parity, only the chiral symmetry-breaking part contributes.  
\begin{equation} 
\begin{split} 
\langle \bar{q} q \rangle &= - \lim_{x \rightarrow 0}  \langle {\rm Tr} [S_q^B(x,0)]
\rangle \\
& = -\pi \langle \rho(0) \rangle, 
\label{bc-rel}
\end{split} 
\end{equation}
where $\rho(0)$ is the zero mode density, a formula derived by Casher and Banks \cite{BC}.

Let us now move on to a typical four-quark condensate.  Such operators have two distinct forms when using quark connectivity: these are the quark-disconnected $(dis)$ and quark-connected $(con)$ pieces denoted below by the respective subscripts,  
\begin{equation} 
\begin{split} 
\langle (\bar{q} \Gamma q)(\bar{q} \Gamma q) \rangle  &= \langle {\rm Tr} [S_{q}^i\Gamma]{\rm Tr}[S_{q}^i\Gamma] \rangle_{dis} - \langle {\rm Tr} [\Gamma S_{q}^i \Gamma S_{q}^i] \rangle_{con},
\label{4-q_sep}
\end{split} 
\end{equation}
where $\Gamma$ is any combination of Dirac, color and/or flavor matrix. The cross terms in the summation in $i=B,S$ do not contribute for the $\Gamma$ matrices that we discuss. 
$S_q^B$ is the difference between the original quark operator and its chiral rotated form: This part does not vanish if the chiral symmetry remains broken. Also,  as given in Eq.~\eqref{bc-rel}, it is proportional to the zero mode density.

Let us look at a few examples.
For now, let us introduce  the flavor matrix   normalized as ${\rm Tr} \frac{\tau^a}{2}\frac{\tau^b}{2}=\frac{\delta^{ab}}{2}$,

\begin{itemize}

\item When $\Gamma=\gamma^\mu$,  using the trace property one finds 
    \begin{equation} 
\begin{split} 
\langle (\bar{q}\frac{\tau^3}{\sqrt{2}} \gamma^\mu  q)(\bar{q}\frac{\tau^3}{\sqrt{2}} \gamma^\mu q) \rangle  &=  - \langle {\rm Tr} [ \gamma^\mu S_{q}^B \gamma^\mu S_{q}^B] \rangle_{con}- \langle {\rm Tr} [\gamma^\mu S_{q}^S \gamma^\mu S_{q}^S] \rangle_{con},
\label{4qv}
\end{split} 
\end{equation} 
with no cross-term between the symmetric and breaking part of the propagator. Hence, it contains both the breaking and symmetric operators.  

On the other hand, one notes
\begin{eqnarray}
\gamma^5 S_{q}^S, \gamma^5=-S_{q}^S , ~~~~~ \gamma^5 S_{q}^B \gamma^5=S_{q}^B
\end{eqnarray}
Hence, 
 to\begin{equation} 
\begin{split} 
\langle (\bar{q}\frac{\tau^3}{\sqrt{2}}  i \gamma^5 \gamma^\mu  q)(\bar{q}\frac{\tau^3}{\sqrt{2}} i \gamma^5 \gamma^\mu q) \rangle  &=  - \langle {\rm Tr} [ \gamma^\mu S_{q}^B \gamma^\mu S_{q}^B] \rangle_{con}+ \langle {\rm Tr} [\gamma^\mu S_{q}^S \gamma^\mu S_{q}^S] \rangle_{con}.
\label{4qa}
\end{split} 
\end{equation} 

Therefore, one can isolate the breaking and symmetric parts as follows
\begin{equation} 
\begin{split} 
\langle (\bar{q}\frac{\tau^3}{\sqrt{2}} \gamma^\mu  q)(\bar{q}\frac{\tau^3}{\sqrt{2}} \gamma^\mu q) \rangle_B &= 
\frac{1}{2} \bigg( \langle (\bar{q}\frac{\tau^3}{\sqrt{2}} \gamma^\mu  q)(\bar{q}\frac{\tau^3}{\sqrt{2}} \gamma^\mu q) \rangle +
\langle (\bar{q}\frac{\tau^3}{\sqrt{2}}  i \gamma^5 \gamma^\mu  q)(\bar{q}\frac{\tau^3}{\sqrt{2}} i \gamma^5 \gamma^\mu q) \rangle \bigg), \\
\langle (\bar{q}\frac{\tau^3}{\sqrt{2}} \gamma^\mu  q)(\bar{q}\frac{\tau^3}{\sqrt{2}} \gamma^\mu q) \rangle_S &=
\frac{1}{2} \bigg( \langle (\bar{q}\frac{\tau^3}{\sqrt{2}} \gamma^\mu  q)(\bar{q}\frac{\tau^3}{\sqrt{2}} \gamma^\mu q) \rangle -
\langle (\bar{q}\frac{\tau^3}{\sqrt{2}}  i \gamma^5 \gamma^\mu  q)(\bar{q}\frac{\tau^3}{\sqrt{2}} i \gamma^5 \gamma^\mu q) \rangle \bigg),
\label{4qv-bs}
\end{split} 
\end{equation}
where the subscript $B,S$
respectively indicate the breaking and symmetric part of the four-quark operator, respectively.  

\item When $\Gamma=1$, we also have the quark disconnect contribution denoted by the subscript $dis$ below.  
    \begin{equation} 
\begin{split} 
\langle (\bar{q} q)(\bar{q} q) \rangle  &= \langle {\rm Tr} [S_{q}^B]{\rm Tr}[S_{q}^B ] \rangle_{dis} - \langle {\rm Tr} [S_{q}^B  S_{q}^B] \rangle_{con}- \langle {\rm Tr} [ S_{q}^S  S_{q}^S] \rangle_{con}.
\label{4qs}
\end{split} 
\end{equation}
To isolate the connected piece above, we introduce the following two quark operators.
\begin{equation} 
\begin{split} 
\langle (\bar{q} i \gamma^5 \frac{\tau^3}{\sqrt{2}} q)(\bar{q} i \gamma^5 \frac{\tau^3}{\sqrt{2}} q) \rangle  &=   \langle {\rm Tr} [S_{q}^B  S_{q}^B] \rangle_{con}- \langle {\rm Tr} [ S_{q}^S  S_{q}^S] \rangle_{con}, \\
\langle (\bar{q} \frac{\tau^3}{\sqrt{2}} q)(\bar{q}  \frac{\tau^3}{\sqrt{2}} q) \rangle  &=  - \langle {\rm Tr} [S_{q}^B  S_{q}^B] \rangle_{con}- \langle {\rm Tr} [ S_{q}^S  S_{q}^S] \rangle_{con}.
\label{4qis}
\end{split} 
\end{equation}
Therefore
\begin{equation} 
\begin{split} 
\langle (\bar{q} q)(\bar{q} q) \rangle_B  &= 
\langle (\bar{q} q)(\bar{q} q) \rangle
-\frac{1}{2} \bigg(
\langle (\bar{q} i \gamma^5 \frac{\tau^3}{\sqrt{2}} q)(\bar{q} i \gamma^5 \frac{\tau^3}{\sqrt{2}} q) \rangle  +
\langle (\bar{q} \frac{\tau^3}{\sqrt{2}} q)(\bar{q}  \frac{\tau^3}{\sqrt{2}} q) \rangle \bigg) 
\\
\langle (\bar{q} q)(\bar{q} q) \rangle_S  &= 
\frac{1}{2} \bigg(
\langle (\bar{q} i \gamma^5 \frac{\tau^3}{\sqrt{2}} q)(\bar{q} i \gamma^5 \frac{\tau^3}{\sqrt{2}} q) \rangle  +
\langle (\bar{q} \frac{\tau^3}{\sqrt{2}} q)(\bar{q}  \frac{\tau^3}{\sqrt{2}} q) \rangle \bigg).
\label{4qsb}
\end{split} 
\end{equation}
Throughout this work, we will use this method: although a similar separation can be made by writing the quark operators in terms of right and left-handed quarks. 

\item The method can be generalized for all $\Gamma$ matrices. We refer to references \cite{Kim:2020zae,Kim:2021xyp} for details.  

\item Because the static heavy quark probes all the gluon configurations, it can be shown that the gluon condensate can also be expressed in terms of the eigenvalues of the Dirac modes\cite{Lee:2013es}. 
\begin{align}
    \langle \frac{\alpha_s}{\pi} G^2 \rangle =12 \langle \sum_\lambda \rho(\lambda) \rangle.
\end{align}
As can be seen from the above formula, the chiral-symmetry breaking effect $\rho(0)$ is multiplied by $d \lambda$ and does not contribute to the gluon condensate.  This shows that the physics of the gluon condensate and that of the chiral-symmetry breaking have different origins.

\end{itemize}

\section{Vector and axial-vector meson mass}

We will now look at how the method can be applied to calculate the hadron masses in the chiral symmetry-restored vacuum.

\subsection{$\rho$ and $a_{1}$}

Let us start by looking at the OPE of the following correlation function
\begin{eqnarray}
\Pi_{\mu \nu}(q) =
i \int d^4x e^{iqx} \langle
T [J_\mu(x), J_{\nu}(0)],
\end{eqnarray}
where $J_\mu=\bar{q}\gamma_\mu \tau^3 q$ for the $\rho $
meson; for the axial vector meson, we extract the spin 1 part of $J^A_\mu=\bar{q} \gamma^5 \gamma_\mu \tau^3 q$.  We use the OPE appearing in the  
polarization function defined as  $\Pi=\Pi^\mu_\mu/(-3q^2)$.  The most important contribution to the vector mesons comes from the dimension-6 four-quark operators. 
These operators appear as 
$-\pi \alpha_s \mathcal{M}/Q^6$, where for $\rho$ and $a_1$ mesons are given as follows, respectively. 
\begin{equation} 
\begin{split}
 \mathcal{M}_{\rho} =&\; 2 \langle (\bar{q} \gamma_\mu \gamma^5 \lambda^a \tau^{3}q)^{2} \rangle  + \frac{4}{9}\langle(\bar{q}\gamma_\mu \lambda^a q)(\sum_{q = u,d,s}\bar{q}\gamma_\mu \lambda^a q) \rangle , \\
  \mathcal{M}_{a_{1}} =&\;  2  \langle (\bar{q} \gamma_\mu \lambda^a \tau^{3}q)^{2} \rangle  + \frac{4 }{9}\langle (\bar{q}\gamma_\mu  \lambda^a q)(\sum_{q = u,d,s}\bar{q}\gamma_\mu \lambda^a q) \rangle. \label{rho-a1-6}
\end{split}
\end{equation} 
Here $\lambda^a$ and $\tau^3$ represent the color and SU(2) flavor matrices, respectively. Hence, $q$ denotes the contribution from both the $u$ and $d$ quarks. On the other hand, the second set of operators has a summation included as the contribution from strange quark is added. We can now use the methods discussed in the previous section and divide the operators into chiral symmetric and breaking pieces. We further introduce the auxiliary parameters $\kappa_{\rho}$ and $\kappa_{a_{1}}$: these parameters reduce to 1 when the vacuum saturation hypothesis is used for the quark-quark operators. 
The four-quark operators can now be written as below. 
\begin{equation} 
\begin{split} 
 \mathcal{M}_{\rho} = & ~
\kappa_{\rho}  \frac{448}{81} \langle \bar{u}u \rangle^{2} \ =\; \frac{28}{9} \langle B_{uu} \rangle_{B}+ \langle S_{\rho-a_{1}} \rangle_{S},\\
 \mathcal{M}_{a_{1}} = & ~
-\kappa_{a_{1}}\frac{704}{81}\langle \bar{u}u\rangle^{2}=\; -\frac{44}{9}\langle B_{uu} \rangle_{B}+ \langle S_{\rho-a_{1}}\rangle_{S},
\end{split}  \label{kappa}
\end{equation}
where 
\begin{equation} 
\begin{split} 
\langle B_{uu} \rangle_{B} =& \frac{1}{2}\bigg(\langle (\bar{u}\gamma_\mu \gamma^5 \lambda^a d)(\bar{d} \gamma_\mu \gamma^5 \lambda^a u) \rangle - \langle (\bar{u} \gamma_\mu \lambda^a d )( \bar{d} \gamma_\mu \lambda^au) \rangle \bigg)_{B} , \\
\langle S_{\rho-a_1} \rangle_{S}  =& \frac{11}{9}\bigg(\langle \bar{q} \gamma_\mu \gamma^5 \lambda^a \tau^{3} q)^{2} \rangle +\langle (\bar{q} \gamma_\mu  \lambda^a \tau^{3} q)^{2}\rangle \bigg)_{S} +\frac{4}{9}\bigg( \langle (\bar{q}\gamma_\mu \lambda^a q)^{2} \rangle- \langle (\bar{q} \gamma_\mu \lambda^a \tau^{3}q)^{2} \rangle \bigg)_{S}\\
&+\frac{4}{9}\langle (\bar{q} \gamma_\mu  \lambda^a q)(\bar{s} \gamma_\mu \lambda^a s) \rangle_{S}.
\end{split} 
\end{equation}
There is an important point to note in the second set of terms in Eq. \eqref{rho-a1-6}.  These terms appear with the same coefficients in both the $\rho$ and $a_1$ parts.  However, the operator has chiral symmetry-breaking contributions: the zero modes contribute to the connected quark diagrams of this operator.

One notes that the symmetric parts of the $\rho$ and $a_1$ channels are identical while the breaking operator $B_{uu}$ contributes with different coefficients.  
Using the observed masses and widths of the $\rho$ and $a_1$ meson in their respective sum rules, we can uniquely determine the vacuum expectation values of the 4-quark operators and thus their breaking and symmetric parts.   Then, one can uniquely determine the respective $\kappa$ values. Using these results, we can estimate the masses of the hadrons in the chiral symmetry-restored vacuum.  This is accomplished for the $\rho$ and $a_1$ mesons by deleting the breaking part of the operator keeping only the chiral symmetric operators to their vacuum value and performing the QCD sum rules. Table \ref{tabkappa} shows the results for both the matrix elements and the masses in the vacuum where chiral symmetry is restored.   
For $\rho-a_1$, the result in Table \ref{tabkappa} is obtained by taking the $a_1$ width to be 400  MeV.  
When $\Gamma_{a_{1}}= 250 (600)$  MeV is taken, we find that $m_{sym}^{\rho-a_{1}}$ decreases (increases) by  about 27.5 MeV.

\begingroup
\setlength{\tabcolsep}{12pt} 
\renewcommand{\arraystretch}{1.5} 
\begin{table}[htbp]
\begin{tabular}{cccc}
\hline \hline
Particle & $\kappa(\sqrt{s_0}({\rm GeV}))$ & $S/B$ & $\bar{m}_{sym}$(MeV)\\
\hline \hline
$\rho$  & 2.60(1.17) & 0.760 & \multirow{2}{*}{572.5 $\pm$ 27.5}\\
$a_{1}$ & 0.76(1.55) & -0.485 \\ \hline
$\omega$& 3.20(1.15) & 1.165 & 655 $\pm$ 15\\
$f_{1}$ & 1.85(1.58) & 0.253 & 1060 $\pm$ 30\\ \hline
$K^{*} $& 2.097(1.33) & 2.831&  \multirow{2}{*}{545 $\mp$ 5}\\
$K_{1}$ & 0.39(1.56) & -0.227&  \\
\hline \hline
\end{tabular}
\caption{The $\kappa$'s (equivalent to Eq. \eqref{kappa} for the respective particles) are evaluated using QCD sum rules with the observed mass and width.  The value in the bracket shows the threshold parameter $s_{0}$.  The ratio $S/B$ shows the fraction of the chiral symmetric part $S$ to the chiral symmetry-breaking part $B$. The mass in the chiral symmetry-restored vacuum is given in the column $\bar{m}_{sym}$. The uncertainties in  $\bar{m}_{sym}$ come from taking the width of $a_{1}$ meson from 250 MeV to 600 MeV,  The central values are those obtained with the $a_{1}$ width of 400 MeV\cite{Kim:2021xyp}.}
\label{tabkappa}
\end{table}

\subsection{Other hadrons}

As with ($\rho$, $a_1$), ($K_{1}$, $K^{*}$) are also chiral partners. On the other hand, ($\omega$, $f_{1}$) do not form chiral partners: this is due to the disconnected diagrams. 
These features can be understood from the OPE perspective. 
Nevertheless, we can start from the symmetry-breaking operators determined from the $\rho$ and the $ a_1$ sum rule and use the physical mass and width of the other hadrons to determine the additional 4-quark operators that are symmetric and breaking chiral under chiral symmetry that appear in the isospin singlet sum rules.  One can then determine the corresponding mass in the chiral symmetric limit.  The details are given in Ref. \cite{Kim:2021xyp}  and summarized in Table \ref{tabkappa}.

\section{Other effects in medium}

In the previous section, we have shown how to estimate the masses of hadrons in a hypothetical vacuum where the chiral symmetry is restored.  Experiments probe finite temperature and/or density configuration where chiral symmetry is restored.  However, in these configurations, there are other effects that smear the signal.   We will discuss other important effects not related to chiral symmetry restoration that affect the mass of a hadron in the medium.

\subsection{Dispersion effects}

The medium can be characterized by its frame of reference, which is usually taken to be at rest $n^\mu=(1,0,0,0)$.  The medium introduces effects that break Lorentz invariance but are not related to chiral symmetry.  In general,  the introduction of $n^\mu$ makes the correlation function in the momentum space a function of both $q^2$ and $q \cdot n$.  Depending on the current and the medium, the polarization can be even or odd with respect to $q \cdot n$.

\subsubsection{Scalar particle}

Let us start with a simple example and consider a charge-neutral scalar particle in an isospin-symmetric medium.  The self-energy $\Sigma(\omega,\vec{q})$ of the scalar particle can be expanded near the mass shell as follows.
\begin{eqnarray}
\Sigma(\omega,\vec{q})=a(q^2-m^2) +b \vec{q}^2+S. 
\end{eqnarray}
Substituting this into the dispersion relation, $q^2-m^2=\Sigma(\omega,\vec{q})$, will lead to pole mass shift that depends on the momentum of the particle
\begin{equation}
    \Delta m^2=\frac{b \vec{q}^2+S}{1-a}.  
\end{equation}
The question relating the mass shift in medium and chiral symmetry breaking is where the effects of the latter are hidden: $a,b$ or $S$.  It is in $S$.  

\subsubsection{Nucleon in medium}

For spin 1/2 particle, there is the additional dependence coming from the gamma matrix.  To understand the complexity, we just assume a self-energy typically parameterized in Walecka type model in nuclear matter at rest.
\begin{eqnarray}
    \slash \hspace*{-0.2cm} q -m =S+\gamma^0 V.
    \label{nucl-phen}
\end{eqnarray}
This leads to the following pole mass shift at $\vec{q}=0$.
\begin{eqnarray}
    \Delta m= S+V  .
\end{eqnarray}  
It is known phenomenologically that there 
are a large scalar attraction ($\sim -400$ MeV) and a vector repulsion of similar magnitude due to the nuclear density; both have small energy dependencies. 

When analyzing the OPE for the nucleon, one introduces a nucleon interpolating field $\eta$ composed of three quarks and studies the two-point correlation function.  
\begin{align}
    \Pi_N(\omega, \vec{q}) = i \int d^4x e^{iqx} \langle T[ \eta(x) \bar{\eta}(0)] \rangle.  \label{nucl-corr}
\end{align}
The phenomenological side of Eq.~\eqref{nucl-corr} can be parametrized using Eq.~\eqref{nucl-phen} as
\begin{align}
     \Pi_N(\omega, \vec{q}) =-\lambda_N^2 \frac{(\omega-V)\gamma^0+(m+S)}{(\omega-V)^2-(m+S)^2}+ ,
\end{align}
where $\lambda_N$ is the coupling between the nucleon interpolating field and the nucleon state. 
Performing the OPE for Eq.~\eqref{nucl-corr},  it was shown that the leading operators that contribute to  the scalar and vector self-energy are respectively given as 
\begin{align}
\begin{split}
     m+S & \propto -\langle \bar{q} q \rangle, \\
     V & \propto \langle \bar{q} \gamma^0 q \rangle .
\end{split}
\end{align}
Therefore, while the scalar attraction is partly related to the decrease of the quark condensate, the vector repulsion is related to the quark density effect.  
Hence, while the scalar attraction is related to the condensate, its quenching can not be directly seen in the pole mass due to the presence of baryon (quark) density.  Furthermore,  as the nucleon interpolating field is not an eigenstate of the parity, there will be contributions from the negative parity nucleon state that contributes with an opposite sign in the scalar mass.  Hence, quenching of the chiral condensate in the OPE will induce the vanishing of the mass difference between the parity partner nucleon states but will not tell us anything about the nucleon mass change just by looking at the scalar part of the correlation function.  

\subsubsection{Vector particles}

There are even more complications for the vector particles due to the different responses of the moving vector meson with respect to the medium.  The vector meson self-energy in the medium can in general be written as follows\cite{Lee:1997zta}. 
\begin{align}
    P_{\mu \nu } (q^2-m^2) & =P^T_{\mu \nu } \Pi^T(\omega, \vec{q})+ P^L_{\mu \nu } \Pi^L(\omega, \vec{q}), \label{vec1}
\end{align}
where $P_{\mu \nu }=(q_\mu q_\nu/q^2-g_{\mu \nu}),~  P^T_{ij}=(\delta_{ij}- \vec{q}_i \vec{q}_j/\vec{q}^2) $, where $i,i$ are spatial indices and the remaining polarization parts begin zero,  and $P^L_{\mu \nu} =(q_\mu q_\nu/q^2-g_{\mu \nu} -P_{\mu \nu}^T )$.  
\begin{itemize}

\item When $\vec{q}=0$ then $\Pi^T=\Pi^L=\Pi(\omega)$ and the mass in medium can be obtained from Eq. \eqref{vec1}, which reduces to below.
\begin{equation}
    (\omega^2 -m^2) =\Pi(\omega).
\end{equation}

\item When $\vec{q} \neq 0$ then $\Pi^T \neq \Pi^L$ and the mass in the medium can be obtained from Eq. \eqref{vec1} separately for the transverse and longitudinal modes, which after the projection reduces to below. 
\begin{align}
\begin{cases}
  {\rm Transverse~~ mode:}   &  (q^2 -m^2)  =\Pi^T(\omega, \vec{q})  \cr
         {\rm Longitudinal~~ mode:} &   (q^2 -m^2)  =\Pi^L(\omega, \vec{q})  
\end{cases}
\end{align}

\end{itemize}
By analyzing the OPE for light vector particles, one notes that the chiral symmetry breaking operators constitute an important contribution to $\Pi$ but do not contribute significantly in $\Pi^T-\Pi^L$.  Hence when measuring chiral symmetry-breaking effects from the medium, it is important to measure vector particles with low momenta with respect to the medium.  

One can estimate the importance of the momentum dependence of the mass shift by looking at the OPE term that contributes to this effect.  To leading order in $\alpha_s$, nuclear density $\rho_n$  and $\vec{q}$ the lowest order operator that contributes to the difference is\cite{Friman:1999wu}
\begin{align}
    \Pi^T(\omega,\vec{q})-\Pi^L(\omega,\vec{q}) = \frac{mA_2^{u+d} \vec{q}^2}{\omega^6} \rho_n. 
\end{align}
Here $m$ is the nucleon mass and $A_2^q=2 \int dx x[q(x) + \bar{q}(x)]$, where $q(x)$ is the quark distribution function inside the proton at scale $\mu$ taken to be 1 GeV for this work.  It is found that this contributes to about 2\% correction to the mass at $\vec{q} \sim 0.5 $ GeV at nuclear matter density\cite{Lee:1997zta}.  Hence, it is important to study the mass shift for vector meson traveling with small momentum.  

\section{Chiral partners: $K_1$ and $K^*$ mesons
}

As discussed above, isolating the effects of chiral symmetry restoration from experimental measurements of hadron masses is problematic as there are many different medium effects.  On the other hand, it can be shown from OPE and/or general arguments that the mass difference between chiral partners depends only on chiral symmetry breaking.  Therefore, it is important to first measure the mass differences between chiral partners.  There are several candidates.  
The obvious choice is the pion and the sigma meson. These particles have been analyzed frequently in a medium, where these particles are expected to become degenerate when chiral symmetry is restored.  On the other hand, the sigma meson has a large width decaying into two pions and hence is very difficult to measure. While the decay width into two pions is expected to decrease as the phase space becomes smaller when the sigma meson mass decreases in a dense medium, the decaying pions will interact with the medium smearing any significant observable signals.  Hence, it is important to identify chiral partners with reasonably small vacuum widths.

The masses and widths of the vector mesons are given in table \ref{t2}.  The $\rho$ and $a_1$ are chiral partners: on the other hand, because they both have large widths, previous attempts failed to measure their mass shift.  The small width $\omega$ and $f_1(1285)$ seem to be more accessible experimentally: but they are isospin singlets and thus are not chiral partners.~\cite{Gubler:2016djf}. Nevertheless,  $\omega$ and $f_1(1285)$ are chiral partners when we neglect the flavor-changing disconnected contributions.  Hence, when experiments can measure their masses, it would still be useful as we can use this information to extract the density dependence of flavor-changing disconnected four-quark condensates.  
Finally, the best candidates seem to be the $K^*$ and $K_1$. 
They are chiral partners and both have reasonably small widths.  Furthermore, the 
chiral symmetry-breaking effects are in the ground states as the spectra of excited states are similar. 
Below, we will now discuss  $K^*$ and $K_1$ states.

\begingroup
\begin{center}
\setlength{\tabcolsep}{12pt} 
\renewcommand{\arraystretch}{1.5} 
\begin{table}[htbp]
\begin{tabular}{ccc|ccc}
\hline  \hline
$J^{PC}=1^{--}$ & Mass & Width & $J^{PC}=1^{++}$ & Mass & Width \\
\hline \hline
$\rho$ & 770 & 150 & $a_1$ & 1260 &  250 -600 \\
\hline
$\omega$ & 782 & 8.49 & $f_1$ & 1285 &  24.2 \\
\hline
$\phi$ & 1020 & 4.266 & $f_1$ & 1420 &  54.9 \\
\hline
$K^*(1^-)$ & 892 & 50.3 & $K_1(1^+)$ & 1270 &  90 \\
\hline
$K^*(1^-)$ & 1410 & 236 & $K_1(1^+)$ & 1400 &  174 \\
\hline \hline
\end{tabular}
\caption{Physical parameters of the vector and axial vector mesons. Units are in MeV.}
\label{t2}
\end{table}
\end{center}
\endgroup

\subsection{$K_1$ and $K^*$ correlation functions}

We will now study the correlation functions for the $K^*$ and $K_1$ mesons\cite{Song:2018plu}. The time-ordered current correlation function is  given by
\begin{eqnarray}
\Pi_{\mu \nu}(\omega, {\bf q})= i\int d^4 x e^{iq\cdot x}\langle |T[j_\mu, \bar{j}_\nu ]|\rangle,
\label{correlator}
\end{eqnarray}
where $q^\mu=(\omega, {\bf q})$ and 
\begin{eqnarray}
j_\mu^{K_1^+} =  \bar{s} \gamma_\mu \gamma_5 u &, & ~~~ j_\mu^{K_1^-}  =  \bar{u} \gamma_\mu \gamma_5 s \nonumber \\
j_\mu^{K^{*+}}  =  \bar{s} \gamma_\mu u &, &~~~ j_\mu^{K^{*-}}  =  \bar{u} \gamma_\mu s.
\label{current}
\end{eqnarray}
We will only consider isospin symmetric nuclear matter.  Therefore, the result obtained by interchanging the $u$ quark with the $d$ quark will be the same. 

The polarization functions for both the vector and axial vector are not conserved.  Therefore they will have contributions from the scalar and pseudo-scalar mesons respectively. 
To extract the spin-1 part, we use the following projection. 
\begin{eqnarray}
\frac{1}{3} (q^\mu q^\nu/q^2-g^{\mu \nu} ) \Pi_{\mu \nu}(q) \stackrel{{\bf q} \rightarrow 0}{\longrightarrow} \Pi(\omega, 0). 
\label{correlator1}
\end{eqnarray} 
Also, because the medium is at rest, the particle dispersion relation will have both a transverse and longitudinal polarization component.  As discussed in the previous section,  we will therefore choose the external three-momentum to be zero $q^\mu=(\omega,\bf{0})$.

\subsubsection{4-quark operators in the $K^*-K_1$ sector}

When we choose the currents to be $J^{K^*}_\mu =\bar{u} \gamma_\mu s$ and $J^{K_1}_\mu =\bar{u}  \gamma_\mu \gamma^5 s$, the corresponding  dimension-6  4-quark operators are given below. 
\begin{widetext}
\begin{eqnarray}
\Pi^{K^*} & = & 
\frac{2 \pi \alpha_s}{Q^6} 
\langle ( \bar{u} \gamma_\mu \gamma^5 \lambda^a s )( \bar{s} \gamma_\mu \gamma^5 \lambda^a u ) \rangle
+ \frac{2 \pi \alpha_s}{9Q^6} 
\langle ( \bar{s} \gamma_\mu \lambda^a  s +\bar{u} \gamma_\mu \lambda^a  u )
( \bar{q} \gamma_\mu \lambda^a  q ) \rangle ,
\nonumber \\
\Pi^{K_1} & =  & 
\frac{2 \pi \alpha_s}{Q^6} 
\langle ( \bar{u} \gamma_\mu \lambda^a s )( \bar{s} \gamma_\mu  \lambda^a u ) \rangle
+ \frac{2 \pi \alpha_s}{9Q^6} 
\langle ( \bar{s} \gamma_\mu \lambda^a  s +\bar{u} \gamma_\mu \lambda^a  u )
( \bar{q} \gamma_\mu \lambda^a  q ) \rangle .
\nonumber \\
\label{ks-k1}
\end{eqnarray}
\end{widetext}
These operators can also be decomposed in terms of chiral symmetry breaking and symmetric pieces.

\subsubsection{Weinberg relations for $K_1,K^*$}

We first study the difference between the correlation functions for the vector and axial vector currents.  Up to dimension 6 operators, it has the following form.   
\begin{align}
& -\frac{2}{Q^2} m_s \langle\bar{u}u\rangle_0 +\frac{2 \pi}{Q^4} \alpha_s \bigg( \langle (\bar{u} \gamma_\mu \lambda^a s )( (\bar{s} \gamma_\mu  \lambda^a u ) \rangle - c.p
 \bigg) \nonumber \\
 & - \frac{8 \pi}{3Q^6}q^\mu q^\nu \alpha_s \bigg( \langle (\bar{u} \gamma_\mu \lambda^a s )( (\bar{s} \gamma_\nu  \lambda^a u ) \rangle_{\cal ST} -c.p.
 \bigg). \label{order}
\end{align}
Here $c.p$ is the operator where the $\gamma_\mu$ is replaced by $\gamma_\mu \gamma_5$ in the 4-quark operator.
The operators appearing in the second line of Eq.~(\ref{order}) are the twist-4 matrix element.  Since the difference between the chiral partners is an order parameter of the chiral symmetry breaking, Eq.~(\ref{order}) is also a chiral order parameter. 
In fact, all the operators to higher dimensions are all order parameters of the chiral symmetry.  When the vacuum saturation hypothesis is applied the operators appearing in Eq.~(\ref{order}) are $m_s \langle \bar{u} u \rangle$ at dimension 4 and $\langle \bar{s} s \rangle \langle \bar{u} u \rangle$ at dimension 6. As one can see, in this case, the operators are chiral symmetry-breaking parts in the strangeness sector multiplied by those in the light quark sector. 
Therefore, irrespective of whether the vacuum saturation hypothesis is valid, the fact that it should be proportional to the chiral order parameter remains valid.  Therefore, to estimate their changes in the nuclear medium, one can approximate their values in terms of the changes in the light and strange condensates.  

To obtain the Weinberg type sum rule in the current case, we follow the same set of approximations as in the original works\cite{Weinberg:1967kj}. That is, for the phenomenological side, we take the vector and axial vector ground state to be different, but take the form for the excited states including the continuum to be the same for both cases.  
We then make an asymptotic expansion in $1/Q^2$ and equate the phenomenological side to the OPE.  Since we work up to dimension 6 operators, we obtain the following two relations. 
\begin{align}
& f_{K^*}^2 m_{K^*}^4-f_{K_1}^2 m_{K_1}^4 = - 2 m_s \langle \bar{u}u \rangle, \nonumber \\
& f_{K^*}^2 m_{K^*}^6-f_{K_1}^2 m_{K_1}^6 = - \frac{64}{9} \pi \alpha_s \langle \bar{u}u \rangle  \langle \bar{s}s \rangle.
\label{Weinberg2}
\end{align}
When the two relations are combined, we obtain the following relation.  
\begin{align}
& f_{K^*}^2 m_{K^*}^4 ( m_{K_1}^2-m_{K^*}^2)  =  -2 m_s \langle \bar{u}u \rangle m_{K_1}^2 + \frac{64}{9} \pi \alpha_s \langle \bar{u}u \rangle  \langle \bar{s}s \rangle. 
\label{Weinberg3}
\end{align}
Hence, we obtain  $m_{K_1}^2=m_{K^*}^2$  when chiral symmetry is restored.  We expect that the chiral order parameter changes by $30 \%$ in nuclear matter, which means that there will be a non-trivial change in the mass difference between $m_{K_1}$ and $m_{K^*}$.  
In a detailed QCD sum rule analysis in Ref.~\cite{Song:2018plu}, it was found that the mass shift in the nuclear matter will be maximally -249 (-35) MeV for  $K_1^-$ ($K_1^+$).

\section{Phenomenological observations}

The decay channel that is dominant for the $K^*$ and $K_1$ mesons are given in table \ref{t1}.
Using the coupling, we find that both $K^*$ ($K_1$) can be produced by a kaon beam via the $\pi$ ($\rho$) exchange with a nucleon.

\begingroup
\begin{center}
\setlength{\tabcolsep}{12pt} 
\renewcommand{\arraystretch}{1.5} 
\begin{table}[htbp]
\begin{tabular}{cc|cc}
\hline \hline
$1^{-}$ & Decay Mode &   $1^{+}$ & Decay mode  \\
\hline
$K^*(892)$ & $K \pi$ (100\%) & $K_1(1270)$ & $K \rho$ (42\%) \\
 & &    &  $K^* \pi$ (16\%) \\
\hline \hline
\end{tabular}
\caption{Dominant hadronic decay channels of $K^*$ and $K_1$ meson.}
\label{t1}
\end{table}
\end{center}
\endgroup

There are different charge states for the $K^*$ and $K_1$. 
The chiral partners are between the same charge states. Therefore when a $K^-$ beam is used on a nuclear target at the JPARC facility, for example, the produced these particles and their final states for $K_1$ will be as follows.  
\begin{align}
K_1^- \to \begin{cases}
\rho^0 K^-  \cr
\rho^- \bar{K}^0  \cr  
\pi^0 K^{*-}  \cr
\pi^- \bar{K}^{* 0}  
\end{cases}
, ~~~~
\bar{K}_1^0 \to \begin{cases} 
\rho^+ K^-  \cr
\rho^0 \bar{K}^0 \cr   
\pi^+ K^{*-}  \cr
\pi^0 \bar{K}^{* 0}  \end{cases}, \nonumber 
\end{align}
and for $K^*$  by 
\begin{align}
K^{*-} \to \begin{cases} 
\pi^0 K^{-}  \cr
\pi^- \bar{K}^{ 0} \end{cases}  
, ~~~~
\bar{K}^{*0} \to \begin{cases} 
\pi^+ K^{-}  \cr
\pi^0 \bar{K}^{ 0} \end{cases} .
\nonumber
\end{align}
Similar production of both $K^*,K_1$ can be achieved by a pion beam at GSI\cite{Paryev:2020ivs}. 

The degeneracy between  $K^*$ and $K_1$ mass when chiral symmetry is restored can also be probed in a relativistic heavy ion collision\cite{Sung:2021myr}. 
This is so because, at the initial stages of the collision, the hadronic phase will undergo a phase transition to the quark-gluon plasma state, where chiral symmetry is expected to be restored. As the system cools down, one will cross the hadronization point where hadrons are formed: statistical hadronization model (SHM) predicts the hadronization temperature at LHC energies to be around 156 MeV\cite{Stachel:2013zma}. At this point, however, the order parameter is still quite small\cite{Ding:2013lfa}.  As the mass difference depends only on the chiral order parameter, the masses between chiral partners will be similar.  
It is well-known that particle production follows the statistical hadronization model (SHM).  
The prediction for the individual particles will depend on the details of the model that is determined by the system size, strangeness enhancement, and whether the system should be treated as a canonical or grand canonical ensemble.  
However, $K^*$ and $K_1$ have the same number of strangeness and their production ratio will be similar when their masses become degenerate. 
Hence we expect that the initial number of $K^*$ and $K_1$ at the chemical hadronization point will be similar.  One still has to consider the possible changes in their number as the systems go through the hadronic phase\cite{Cho:2015qca}. 
This effect is called the hadronic rescattering effect, which typically depends on the vacuum width of the hadron of interest.  Fortunately, here again, the small and similar vacuum width between $K_1$ and $K^*$ will lead to a similar reduction so that the ratio can be estimated perturbatively.  
If the vacuum masses are used in the SHM, the expected particle ratio between $K_1$ and $K^*$ will be very small due to the much larger mass of $K_1$ in the vacuum. The anomalously large ratio is more enhanced in a peripheral collision, as the hadronic lifetime will be shorter there. 
Therefore measuring the production of both the $K_1$ and $K^*$ from heavy ion collision in both central and peripheral collisions and then comparing the observed production ratios to those obtained in the SHM with vacuum masses will provide a signature for chiral symmetry restoration in heavy ion collisions.

\section{Conclusions}

Understanding the origin of the masses of hadrons is a key topic of interest.  It is believed that chiral symmetry breaking is partly responsible for the generation of the hadron mass in QCD.  Here, using OPE perspective,  we discussed how one can isolate the effects of chiral symmetry breaking in the hadron mass.  While chiral symmetry is expected to be resorted in the medium, there are difficulties in measuring the mass shift and isolating the effects of chiral symmetry breaking.  
We have used OPE perspective to analyze the other effects.  We have further emphasized that to isolate the chiral symmetry-breaking effects, one has to start by measuring the mass differences between chiral partners.  Finally, we showed that experimentally measuring the mass shift between $K^*$ and $K_1$ mass should be the first priority.  

\vspace{6pt} 




\acknowledgments{I would like to thank Jisu Kim for collaborating on many works cited in the references. This work was supported by Samsung Science and Technology Foundation under Project Number SSTF-BA1901-04.}

\end{document}